**Terahertz photometer to observe solar flares in continuum**

Rogerio Marcon • Pierre Kaufmann • Luis Olavo T. Fernandes • Rodolfo Godoy • Adolfo Marun • Emilio C. Bortolucci, • Maria Beny Zakia • José Alexandre Diniz • Amauri S. Kudaka

R. Marcon
"Gleb Wataghin" Physics Institute, State University of Campinas, Campinas, Brazil
and "Bernard Lyot" Solar Observatory, Campinas, Brazil

P. Kaufmann (corresponding co-author)
Center of Radio Astronomy and Astrophysics, Engineering School, Mackenzie Presbyterian University, São Paulo, Brazil and Center of Semiconductor Components, State University of Campinas, Campinas, Brazil
email: pierrekau@gmail.com
Phone: +55 11 2114 8331
Fax: +55 11 32553123

E.C. Bortolucci • M.B. Zakia • J.A. Diniz
Center of Semiconductor Components, State University of Campinas, Campinas, Brazil

L.O.T. Fernandes • A.S. Kudaka
Center of Radio Astronomy and Astrophysics, Engineering School, Mackenzie Presbyterian University, São Paulo, Brazil.

A. Marun • R. Godoy
El Leoncito Astronomical Complex, San Juan, Argentina.

**Abstract**
Solar observations at sub-THz frequencies detected a new flare spectral component peaking in the THz range, simultaneously with the well known microwaves component, bringing challenging constraints for
interpretation. Higher THz frequencies observations are needed to understand the nature of the mechanisms occurring in flares. A THz photometer system was developed to observe outside the terrestrial atmosphere on stratospheric balloons or satellites, or at exceptionally transparent ground stations. The telescope was designed to observe the whole solar disk detecting small relative changes in input temperature caused by flares at localized positions. A Golay cell detector is preceded by low-pass filters to suppress visible and near IR radiation, a band-pass filter, and a chopper. A prototype was assembled to demonstrate the new concept and the system performance. It can detect temperature variations smaller than 1 K for data sampled at a rate of 10/second, smoothed for intervals larger than 4 seconds. For a 76 mm aperture, this corresponds to small solar burst intensities at THz frequencies. A system with 3 and 7 THz photometers is being built for solar flare observations on board of stratospheric balloon missions.

# 1 Introduction

Recent solar flare observations carried out at higher frequencies (0.2 and 0.4 THz) by the Solar Submillimeter Telescope, SST [1], have fully characterized the sub-THz flux spectral component that increases with frequency simultaneously but spectrally separated from the well known microwave component [2-6]. Figure 1 shows the largest solar burst observed by SST exhibiting the two spectral components, one in the GHz and the other in the sub-THz range of frequencies [2]. Time structures of the order of minutes exhibit peak intensities of 18000 SFU at 0.4 THz and 11000 SFU at 0.2 THz (one SFU = $10^{-22}$ W m$^{-2}$ Hz$^{-1}$). The superimposed sub-second time structures exhibit relative intensities larger than 1000 SFU. One of the smallest burst observed by SST is shown in Figure 2, exhibiting temperature variations on time scales of seconds equivalent to 100 SFU at 0.4 THz, , and less than 10 SFU at 0.2 THz [6]. The time scales for the impulsive phase of bursts may range from a fraction of a second to minutes.

Several suggestions where proposed to interpret the sub-THz spectral component [7], however the double spectral feature cannot be well explained by the existing models. One of the possibilities is a mechanism observed in laboratory accelerators that might become important also in solar flares. In this case the high frequency flare emission is attributed to incoherent synchrotron radiation (ISR) produced by accelerated beams of high energy electrons with intensity peaking at THz frequencies. Moreover certain wave-particle instabilities may set in the electron beam, giving rise to bunching of the electrons which radiate powerful broadband coherent synchrotron radiation (CSR) in the microwave spectrum peaking at wavelengths comparable to the size of the bunching [8]. This mechanism may be present in solar flares [9], where the existence of ultra-relativistic electron beams is suggested by hard X- and gamma-ray photon energy spectra in certain flares [10]. Simulations have shown that the mechanism may be extremely efficient and highly localized in solar flares [11,12]. To fully understand the nature of the high frequency emission in flares it is necessary to measure the complete continuum spectra at higher THz frequencies. This requires observations with detectors placed outside the terrestrial atmosphere, as it has been done at far IR for non-solar experiments on SOFIA high altitude aircraft [13], PACS experiment on HERSCHEL satellite [14], a solar scanning experiment on a stratospheric balloon [15], or through few atmospheric THz transmission "windows" at exceptionally good high altitude ground based locations [16-18].

Detection of excess solar continuum radiation in the THz range poses several technological challenges [19-21]. They include the efficient suppression of the powerful visible and near infra-red background component of solar disc emission, the band-pass filtering and the appropriate selection of uncooled detector system and optical setup design to obtain enough sensitivity and time resolution [22-25].

We present a new design concept for a THz photometer, the laboratory prototype construction, and its performance. Although most of the solar burst models predicts THz fluxes considerably larger than the minimum detected at sub-THz, the system was intended to detect minimum fluxes above about 100 SFU, with time resolution of about 100 ms. It was designed to increase the detection probability of a solar transient on the relatively short flight duration of a stratospheric balloon mission, as well as to be prepared for the possibility that there might be an unpredicted emission component existing solely in the THz range.

## 2 Detection of small angular size bursting sources on the solar disk

Solar flares are produced in active regions which are considerably smaller than the solar disk (less – or much less - than one arc-minute on a 32 arc-minutes diameter disk). The impulsive component occurs in small time scales (a fraction of one second to tens of seconds) at different locations in the solar disk which are not known in advance. Therefore, it is difficult to detect bursts with enough sensitivity without missing data in time or in space. Higher sensitivity requires larger aperture sizes which in turn reduces the diffraction-limited main beam to angles smaller than the solar disk. Predicting the location of the flaring active region is not feasible on a short time scale, while raster scanning the whole solar disk might take longer than the duration of the burst time features. The problem is how to observe the whole Sun with sufficient sensitivity to detect flares subtended by much smaller sizes.

One way to overcome this difficulty is to use an optical setup that combines two well known definitions [26]: (a) Forming the full solar disk image on the detecting device surface at the focal plane according to the relationship

$$f \tan \theta = d \qquad (1)$$

where f is the focal length, $\theta$ the Sun angular diameter and d the size of the solar disc image at the focal plane. This relationship is independent from the aperture diameter and from the wavelength. (b) The gain is proportional to the aperture area. It is possible to have a sufficiently large aperture, maintaining the same focal length given by equation (1), to obtain enough gain to detect a desired lower temperature variation and its equivalent flux.

The solar disk image may become blurred because of aberrations when enlarging too much the aperture, keeping the same focal length. For the present photometry application, however, there are no concerns about the quality of the solar image formed in the focal plane, provided that all photons are concentrated on the sensitive device.

On the other hand, adopting the usual radio astronomy relationships defining intensity of radiation [27], the excess emission flux density produced by a solar flare $\Delta S$ is related to a relative excess temperature variation at the input of the sensitive detector $\Delta T$ according to

$$\Delta S = 2 k \Delta T/A_e \qquad (2)$$

where k is the Boltzmann constant and $A_e$ is the effective aperture area. This measurement does not require the knowledge of the absolute temperature of the background. Therefore a physical size aperture can be assigned to provide a minimum detectable temperature variation $\Delta T$ at the sensing device corresponding to a minimum flux density to be observed, irrespectively from where it originates in the solar disk.

Similar results may be obtained by using non-imaging concentrators, such as the Winston cone [28,29]. The acceptance angle of incoming radiation is set by geometrical relationships between the cone input aperture, length, and the size of the back output aperture over which the radiation is concentrated after multiple bounces inside the cone. However, for acceptance angles of the order of the Sun's diameter, this concept may require unrealistically long cones, difficult machining techniques to produce accurate cone internal surface with little loss, for the large aperture size needed for the necessary gain and for the small sensor detecting area dimensions.

The system developed here presents other advantages. It can be built using easy to make, or commercially available optical components. The angle of acceptance of incoming radiation can be increased by enlarging the sensitive detection surface at the focal plane adding, for example, a small non-imaging concentrating cone in front of the sensor's exposed surface. The increase in the acceptance angle reduces the pointing and tracking accuracy requirements for the instrument.

**3 A prototype THz solar photometer**

A solar flare photometer measures a relative excess temperature $\Delta T$ at the input of the detection device, within time scales characteristic of most impulsive bursts (from a fraction of a second to minutes). The measurements are taken with respect to a certain background level, produced by the solar disc black-body radiation, which brightness temperature at THz is of about 4500 K [30]. The absolute background temperature is not required to be known for the measurement of relative enhancements in temperature caused by solar bursts. However, particular care must be taken to remove all the black-body background radiated power from the solar disk concentrated in the visible and near infrared. This power is reduced by four orders of magnitude for the spectrum at all frequencies smaller than 15 THz ($\lambda > 20\mu$). At discrete THz frequencies (through $\pm$ 10% frequency band-pass) the power further reduces to $10^{-6}$ of the radiation in visible and near IR. Therefore, the photometer developed to measure excess THz radiation, with the solar disk as a background, requires complete suppression of the visible and near IR radiation, and a detector sensitive to radiation enhancement due to small solar flares.

Several different experimental THz radiometer systems were mounted on several occasions, involving four laboratories in these developments. They were intended to test and characterize different types of low-pass filters, band-pass filters and options for detectors at room temperature, using temperature-controlled backbody radiation sources [22-25]. The objective, as mentioned before, was to verify the minimum detectable temperature variations, instrumental time constant, and other operational performance. The best solution found for the suppression of the visible and near-IR radiation was obtained by a combination of a first reflection of incoming radiation by a roughened surface mirror [31,32], followed by TydexBlack low pass membrane [33]. The THz band-pass resonant metal meshes were locally fabricated, exhibiting excellent performance from sub-THz up to 10 THz, with band-pass of about $\pm$ 10% of the central frequencies [34].

Three types of uncooled sensors were tested for THz radiation detection: a 10$\mu$m camera adapted for THz, a pyroelectric modular sensor, and the Golay cell [see details in 22-25]. Two cameras for the 10$\mu$m band were adapted using HRFZ-Si windows on the focal plane array for better THz response. Photometric measurements were made by adding all FPA pixels. This sensor exhibited rapid time response (milliseconds). However, after adding the membrane low-pass filter (allowing only f < 15 THz), the radiation temperature changes were barely detected for the whole THz range. After adding a band-pass THz filter, the system was unable to respond to any temperature variations over the whole range of 300-1000 K. This is not surprising since the array of microbolometers was designed for 10$\mu$m and should be highly ineffective at much longer wavelengths. Same tests were performed for a pyroelectric module. It responded well to temperature variations over the whole THz band (f < 15 THz), considerably better than the microbolometer array. However, with interposed band-pass filters, these variations could not be detected. It exhibited large time constant (300-500 ms) and pronounced instabilities. The Golay cell, of a recently fabricated model [35], presented an adequate performance for detection of temperature variations, with nearly 50 seconds time constant, for the whole THz band (f < 15 THz) as well as at selected THz frequencies.

Two Golay cells were tested and characterized over three years. They were used on the measuring setups mentioned above, installed at different occasions at each one of the four laboratories engaged in these developments (i.e. at CCS-Unicamp, and OSBL in Campinas, Mackenzie in São Paulo, El Leoncito mountain site, Argentina). No particular attention was given to environmental control at the distinct installations. The Golay cells performance has been strikingly repeatable. It has been submitted to many turn-ons, turn-offs, over short or long time intervals, at distinct places. Once turned on, after a stabilizing time of about 20 minutes, the Golay cell output remains stable within the limits tolerable by the present measuring goals, as will be shown below. The performances of the two Golay cells tested were very similar to each other.

Figure 3 shows the complete THz photometer prototype together with an "artificial Sun" calibration source, assembled at CCS/Unicamp for the tests shown here. The Newport Oriel black-body source (1) output windows selection wheel was placed at the focus of a 150 mm concave mirror (2) with 600 mm focal length, reflecting a parallel beam of radiation. The incoming radiation is first reflected by a flat mirror at 45 degrees inclination, with roughened surface (3) to diffuse a substantial fraction of the visible and near IR power [31,32]. The radiation is then directed into a simple commercial Celestron FirstScope 76 mm aperture Newtonian telescope, with 300 mm focal length (4). The radiation reflected by the telescope secondary flat mirror is directed to the Golay cell (8), passing through the resonant tuning fork 20 Hz chopper (5). The cell input is preceded by a metal mesh band-pass filter fabricated at our laboratory (6), with central frequency at 2 THz (±10%) [34] and a TydexBlack low-pass filter membrane (7) [33] for complete suppression of the visible and near IR emissions [22,23]. A black-body window with 5.08 mm (0.2") diameter is seen as a 0.48 degrees disk after reflection in the mirror (2), which corresponds approximately to an "artificial solar disk" for testing purposes. The telescope forms a solar image of about 3 mm at the focal plane where the Golay cell 10 mm cone is placed to concentrate radiation on a 5 mm sensitive surface. Before the final readout, the Golay cell output was amplified, rectified, and RC integrated over 200 ms (9).

The system response to temperature variations and stability has been evaluated from a 23.7 minutes measurements of the prototype continuous response to a constant temperature increase (from 600 to 1300 K). The system output signal from unit (9) shown in Figures 3 (a) and (b), after a 200 ms integration, has been read digitally, at rate of 10 samples per second. The results are shown in Figure 4, with readings at 10 samples/s and no running mean (top), 10 samples running mean (middle), and 100 samples running mean (bottom). The black-body source exhibited a steady temperature increase ramp with constant slope. It can be seen that (a) the noise drops substantially for longer running means, i.e., for a larger number of data points, or for the equivalent length of time; (b) the noise expressed in relative temperature variations remains the same for increasing temperatures; and (c) the system output response with temperature is linear.

There are various methods to demonstrate the stability for the whole assembled system (without controlled ambient environment). The simple inspection of the plots shown in Figure 4 indicates that for larger smoothing (or running means) the fluctuations reduce dramatically. Slow variations of the order 10 K on time scale of several minutes are also observed. The Allan statistics [36], used to characterize frequency stability of oscillators, might be useful to describe quantitatively the system fluctuations. The Allan variance has been derived for the 14200 points measured along 23.7 minutes for different running means, or number of successive data points (N), i.e. $\sigma_N^2 (N) = (1/2) \langle (\Delta T)^2 \rangle$ where $\Delta T = \delta T_n - \delta T_{n+1}$, with $\delta T_n$ and $\delta T_{n+1}$ the successive temperature differences with respect to the best fit straight line for the entire set of 14200 data points. The Allan Deviation (i.e., the square root of Allan variance, expressed in K) is shown in Figure 5 for different number of points used in the running means (or for the equivalent intervals of time). The same results can be obtained keeping the black-body at a fixed temperature. The Allan Deviation for the whole system shows that temperature variations $\Delta T$ smaller than 1 K can be detected when smoothing over more than 40 data points sampled at a 10/s rate (or equivalently over 4 seconds interval). The result is representative of the whole system performance, which includes the data integrated over 200 ms before being sampled at a 10/s rate, and slow varying time features of unknown origin. The latter might include causes external to the detector, such as uncontrolled variations in the ambient environment.

The approximate noise equivalent power (NEP) for the whole system might be estimated. For signal-to-noise ratio equal to one, NEP $\approx \Delta P_S (\tau)^{0.5}$ [37], where $\Delta P_S = 2 k \Delta T \Delta f$ is the noise equivalent power fluctuations measured within a band-pass $\Delta f$, and k the Boltzmann constant. The 2 THz metal mesh filter band pass $\Delta f$ is of about $\pm 10$ percent, or $4 \cdot 10^{11}$ Hz. With integration time of $\tau = 0.2$ s it has been measured $\Delta T \approx 40$ K at the RC integrator output. We obtain NEP $\approx 2 \cdot 10^{-10}$ W Hz$^{-0.5}$. This approximate estimate is twice the nominal Golay cell specification using 20 Hz chopper [35].

The telescope aperture efficiency, defined as the ratio between the effective aperture to the physical aperture, can be estimated approximately. The principal loss is caused by the blockage of incoming radiation by the sub-reflector and supporting arms. Other losses might be caused by minor residual misalignments. We may assume conservatively a 50% aperture efficiency with an uncertainty of $\pm 20\%$. From equation (2), using a 76 mm physical aperture, a minimum detectable temperature $\Delta T \sim 1$ K corresponds to a minimum detectable flux density $\Delta S$ of the order of 100 SFU. This qualitative performance is perfectly adequate for the proposed application. Improved sensitivity for temperature variations are expected for the definitive photometer model, with the use of Cassegrain telescope with rough primary surface and with temperature controlled electronics and sensors in a sealed case.

**4 The 3 and 7 THz photometers for stratospheric balloon platform**

A THz dual frequency photometer system to operate outside the terrestrial atmosphere is currently being built. It has been designed accordingly to the prototype described here to be operated on board of long-duration stratospheric balloon flights. The experiment has been labeled SOLAR-T. It consists essentially of the same blocks shown in Figure 3, duplicated to operate at two frequencies centered at 3 and 7 THz, with modification in the optics and mechanical assembly. Figure 6 shows the schematic diagram for one photometer with the upper panel giving a possible two channel concept configuration. The definitive telescope has a Cassegrain geometry, using standard Edmund Optics 76 mm concave primary mirror with 76 mm focal length, a 25 mm secondary convex mirror with -25.8 mm focal length. The overall effective focal length of the Cassegrain is related to the system magnification, which depends on the secondary distance from the primary (see for example [38]). The two mirrors have been adjusted optically to produce a 500 mm focal length, before the primary reflector has been roughened. This setup produces a solar disk image of 4.4 mm at the system focal plane, smaller than the Golay cell 10 mm photon trap cone placed in front of the sensitive surface. The solar disk image on the focal plane may be displaced by about 5 mm from the center and still have all radiation photons reaching the detector. According to equation (1) the tolerance angular displacement for a focal length of 500 mm is equivalent to an acceptable mispointing of up to 0.6 degrees, through which all photons are detected.

The primary surface has been roughened, using 10μm Carborundum particles producing 1.25μm r.m.s. surface roughness, which is optimum for diffusion of a significant fraction of radiation with $\lambda < 30\mu$ [32].

Two identical 76 mm diameter Cassegrain telescopes have been built at "Bernard Lyot" Solar Observatory machine shop, Campinas, Brazil, shown in Figure 7. Before being sent to be integrated on the final flying model, each telescope has been successfully tested for visible and near-IR diffusion effectiveness, and for their response to temperature changes, using the setup shown in Figure 8. The arrangement is the same as shown in Figure 3(a), except that the telescope is a Cassegrain with rough primary surface facing directly the concave mirror which reflects the "artificial sun" radiation of the black-body source.

## 5 Concluding remarks

The SOLAR-T experiment with the 3 and 7 THz photometers is planned to fly on board of a long-duration stratospheric balloon flight, coupled to the GRIPS gamma-ray experiment [39] in cooperation with University of California, Berkeley, US. One engineering flight is scheduled for fall 2012 in the USA, and a 2 weeks flight over Antarctica in 2013-2014. Another long duration stratospheric balloon flight over Russia (one week) is planned in cooperation with the Lebedev Physics Institute, Moscow (2015-2016).

The two telescopes are being coupled to a temperature and pressure controlled compact case containing the Golay cells, choppers, low-pass and band-pass filters and other auxiliary devices, at Tydex, Saint Petersbourg, Russia. The environment control of the hardware is expected to further improve the photometers' performance, in comparison to the prototype system shown here. Data acquisition and conditioning subsystem and telemetry subsystem are being developed by the Brazilian companies Propertech and Neuron, respectively. Integration of the complete experiment will be carried out by engineers from the three companies. Final integration to the GRIPS experiment will be carried out at the Space Science Laboratory, University of California, Berkeley, in preparation for the engineering test flight scheduled for the fall 2012.

**Acknowledgements.** We acknowledge and thank the reviewers comments that have contributed for the improvement of the paper presentation. The researches on THz materials, filters and detectors have been developed within a cooperation between the Center for Semiconductor Components of Campinas State University, Campinas, SP, Brazil, the "Bernard Lyot" Solar Observatory, Campinas, SP, Brazil, the Center for Radio Astronomy and Astrophysics, Engineering School, Mackenzie Presbyterian University, São Paulo, SP, Brazil, the Complejo Astronomico El Leoncito, Argentina, funded by Brazilian agencies FAPESP, CNPq INCT-NAMITEC, CNPq, and Mackpesquisa, and Argentina agency CONICET. The SOLAR-T THz dual photometers experiment is being funded by FAPESP.

**Captions to the Figures**

Figure 1 – The largest sub-THz solar burst observed by the SST telescope [1] that occurred on November 4, 2003 showing the time profiles at 0.2 and 0.4 THz. The bottom panel shows a 20 s zoom of the first peak of the event, exhibiting intense sub-second time structures. The inner panel shows the "double spectral" features, with the well known microwaves and the new THz component [2,3].

Figure 2 – One of the smallest solar burst recorded on February 8, 2010, showing complex time structures observed at 0.4 THz only. Intensity is given in equivalent antenna excess noise temperature, with a peak of about 150 K at 0.4 THz and less than 10 K at 0.2 K, corresponding to 100 and 4 solar flux units, respectively [6].

Figure 3 – The THz photometer prototype arrangement together with the black-body radiation source used for tests. The diagram in (a) shows the "artificial Sun" setup, with the black-body (1) radiation output window at the focus of a concave mirror (2). Part of the parallel beam coming from the concave mirror is reflected by a flat mirror (3), at 45º inclination, with rough surface, and directed to a 76 mm newtonian telescope (4), concentrating the radiation to a Golay cell detector (8) preceded by a resonant tuning fork chopper (5), a 2 THz resonant metal mesh band-pass filter (6) and a TydeBlack

low-pass membrane filter (7). The cell output is amplified, rectified and integrated (9) for the measurement. A picture of the setup is shown in (b) installed at CCS/Unicamp.

Figure 4 – The 2 THz photometer response to the black body "artificial Sun" temperature continuous heating up, from 600 to 1300 K, with 10 readings per second. The sample shows 14200 data points, throughout 23.7 minutes. The top shows the raw data, after a 200 ms RC analog integration taken from the output of unit (9) shown in Figure 3, sampled at the rate of 10/s, without any smoothing. The plots in the middle and bottom panels correspond to running means of 10 and 100 points, respectively.
Figure 5 – The stability of the system shown in terms of Allan Deviation for the data shown in Figure 4(top), obtained for various running means (in number of points, or in equivalent time intervals containing the respective number of points). Relative temperature variations smaller than 1K can be measured for smoothing with more than 40 data points (or over 4 seconds interval).

Figure 6 – Schematic diagram showing the principal components of the THz photometer (left), using a Cassegrain optical configuration (right). The upper panel shows the dual 3 and 7 THz photometers conceptual configuration for the flying model.

Figure 7 – Two 76 mm diameter Cassegrain telescopes built for the SOLAR-T experiment. The optical alignment was done before the primary mirrors surfaces were roughened.

Figure 8 - Laboratory setup to test the response of the 76 mm Cassegrain telescopes for the "artificial Sun" temperature variations. The picture shows (a): the black-body source (1); output window selection wheel (2), the concave mirror (3), the 76 mm telescope (4), the Golay cell (5), the chopper(6), 2 THz metal mesh band pass filter (7) and TydexBlack low-pass membrane (8).

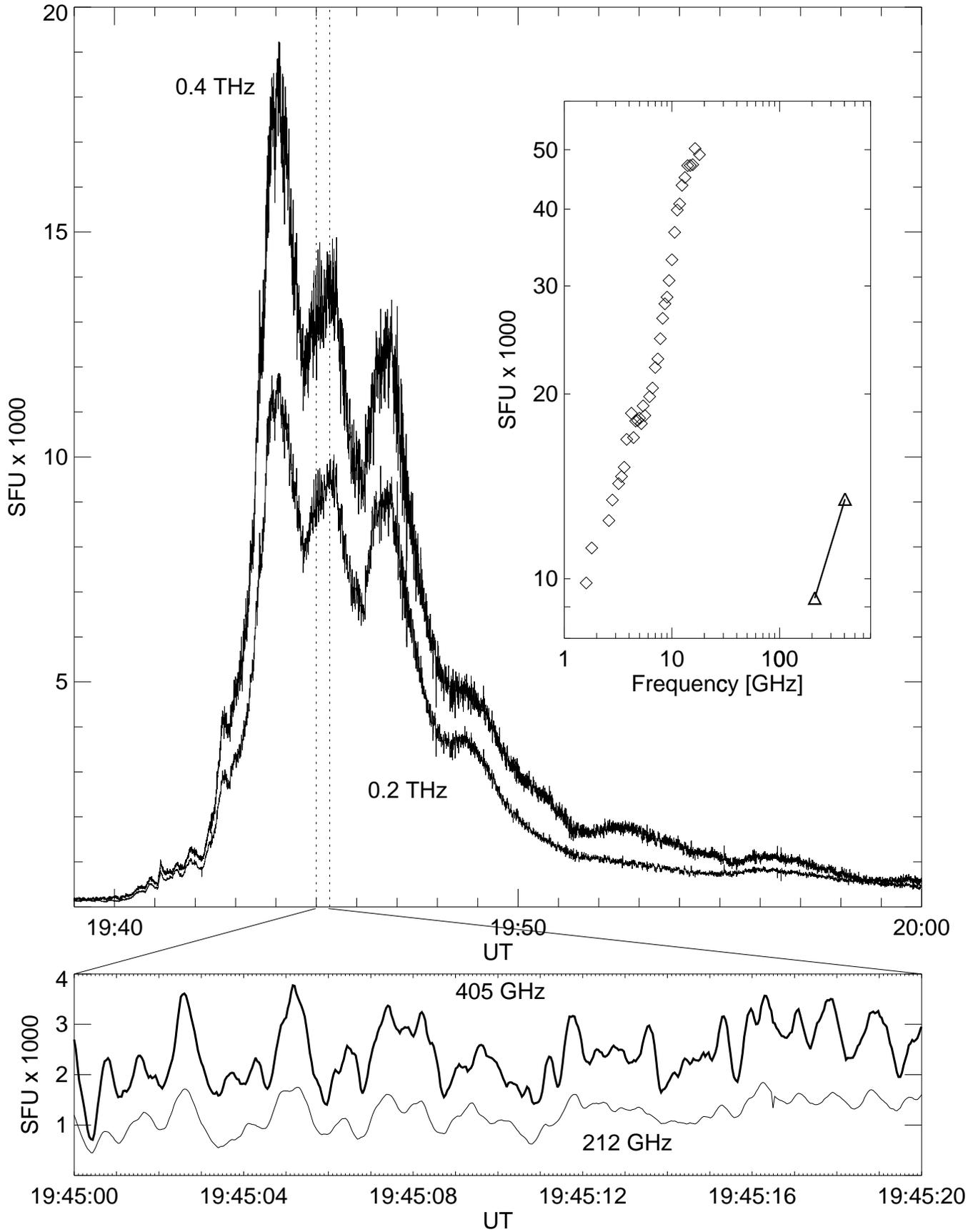

Figure 1

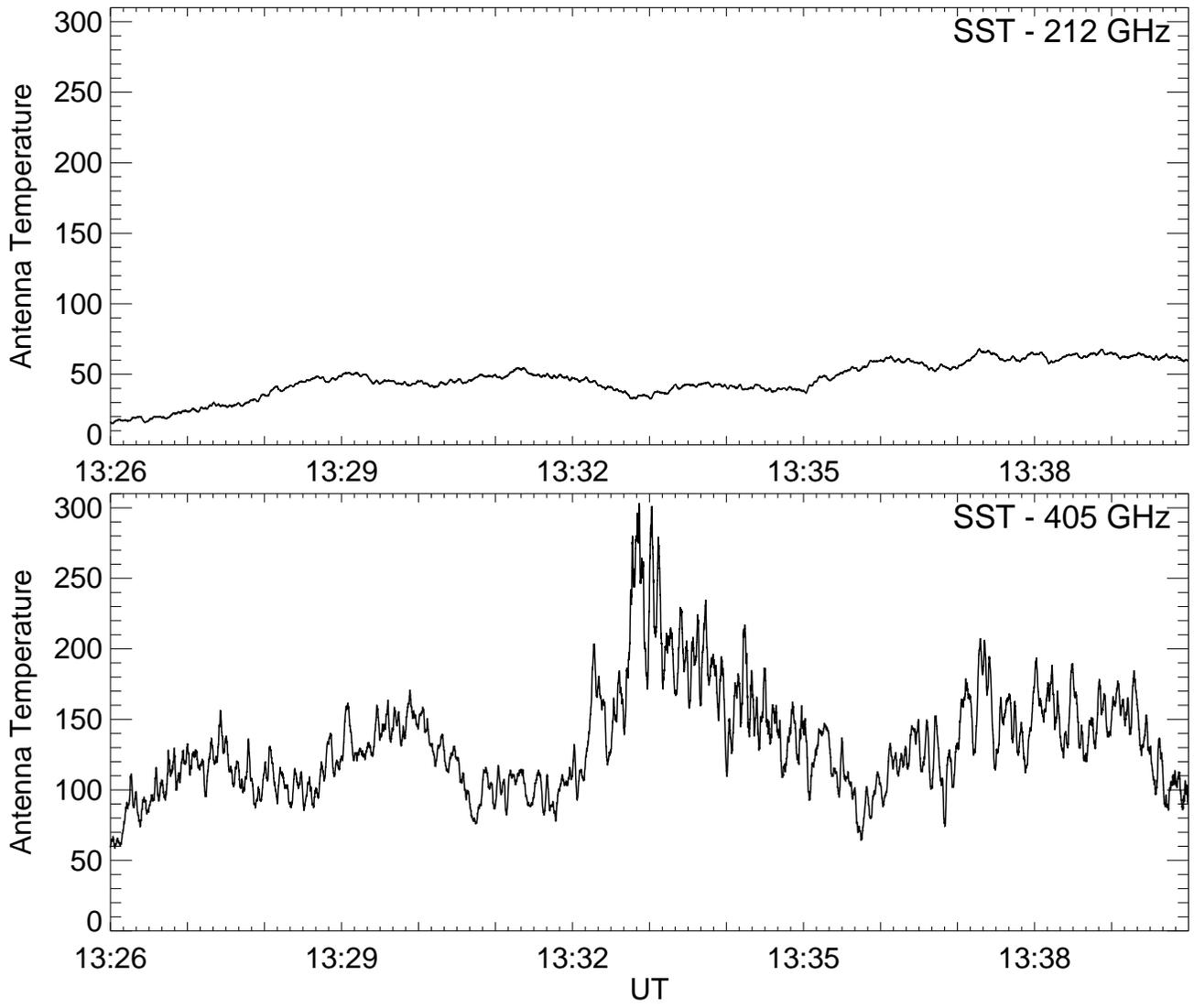

**Figure 2**

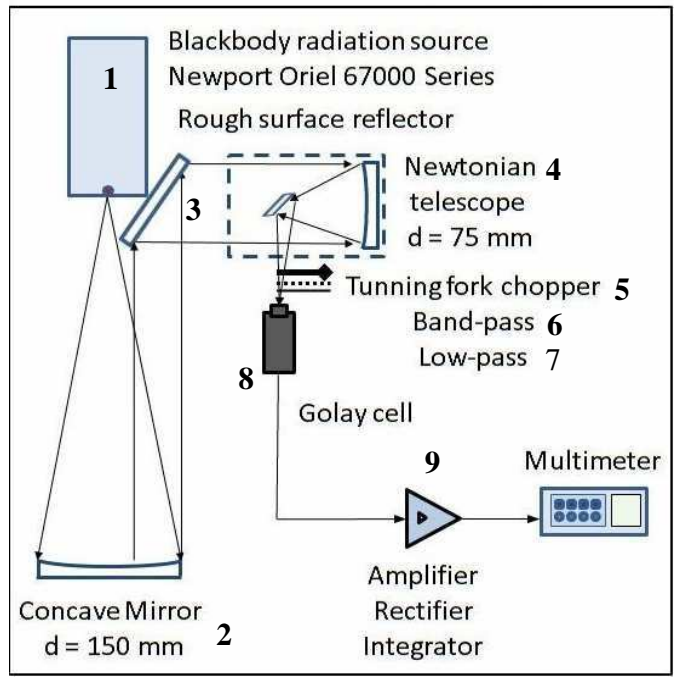 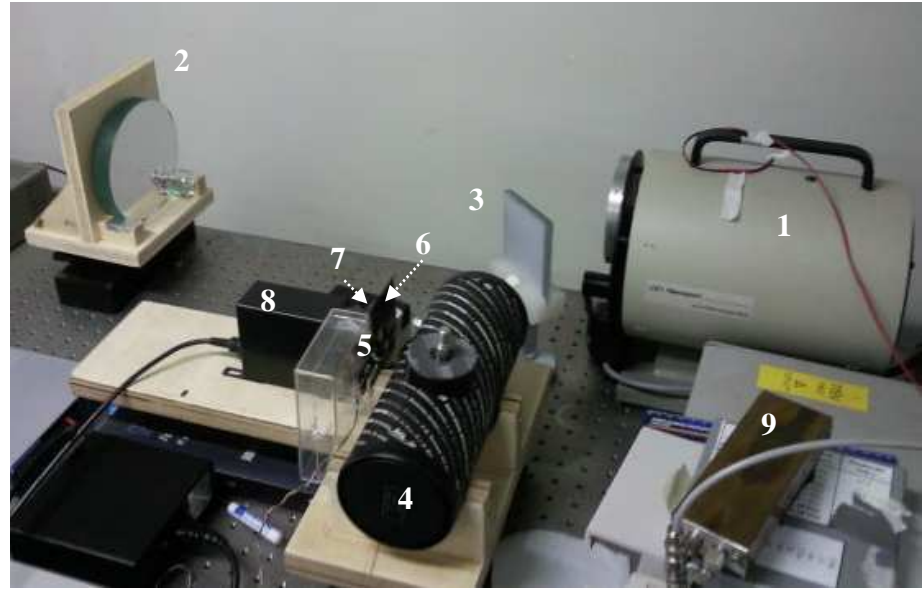

(a) (b)

Figure 3

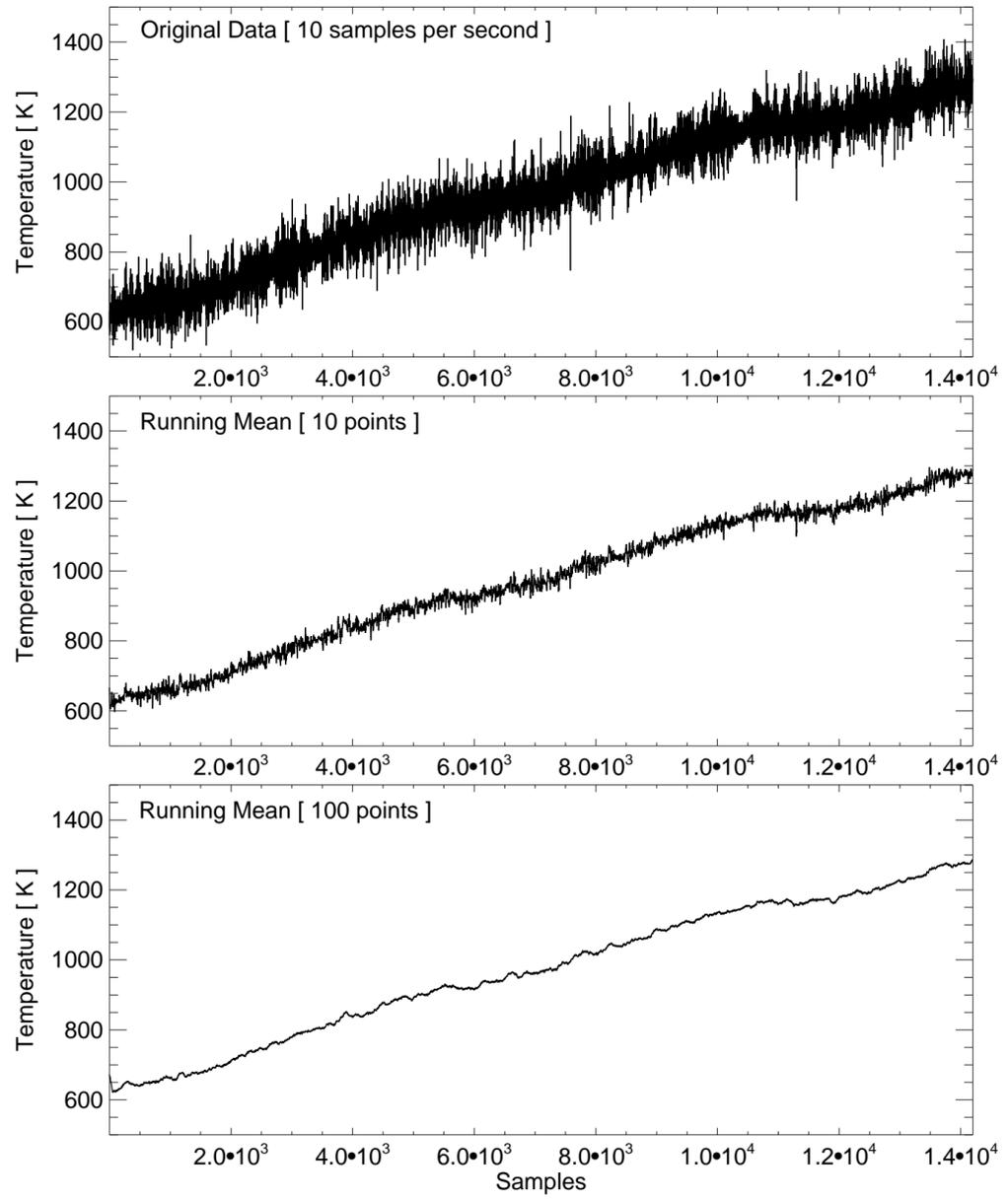

Figure 4

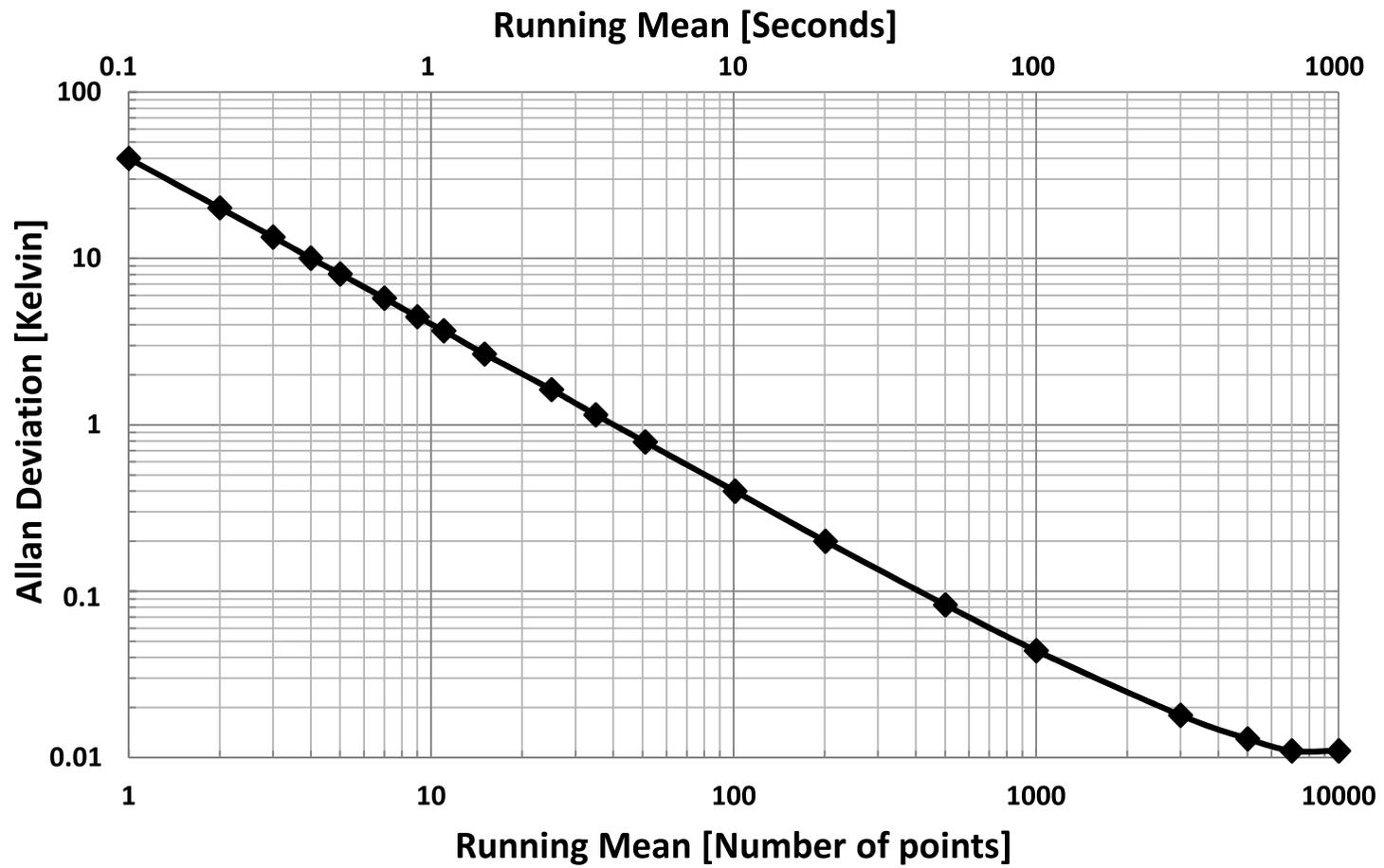

Figure 5

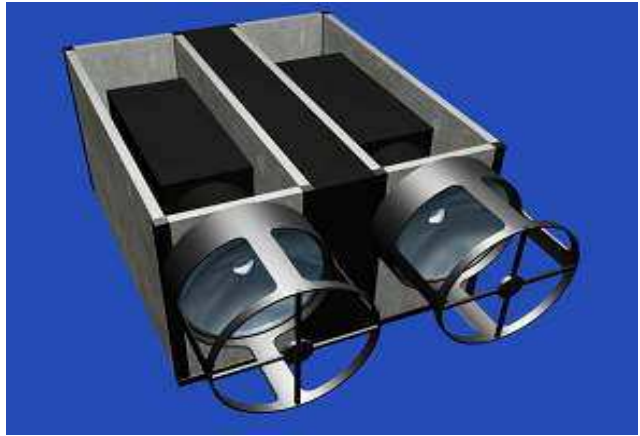

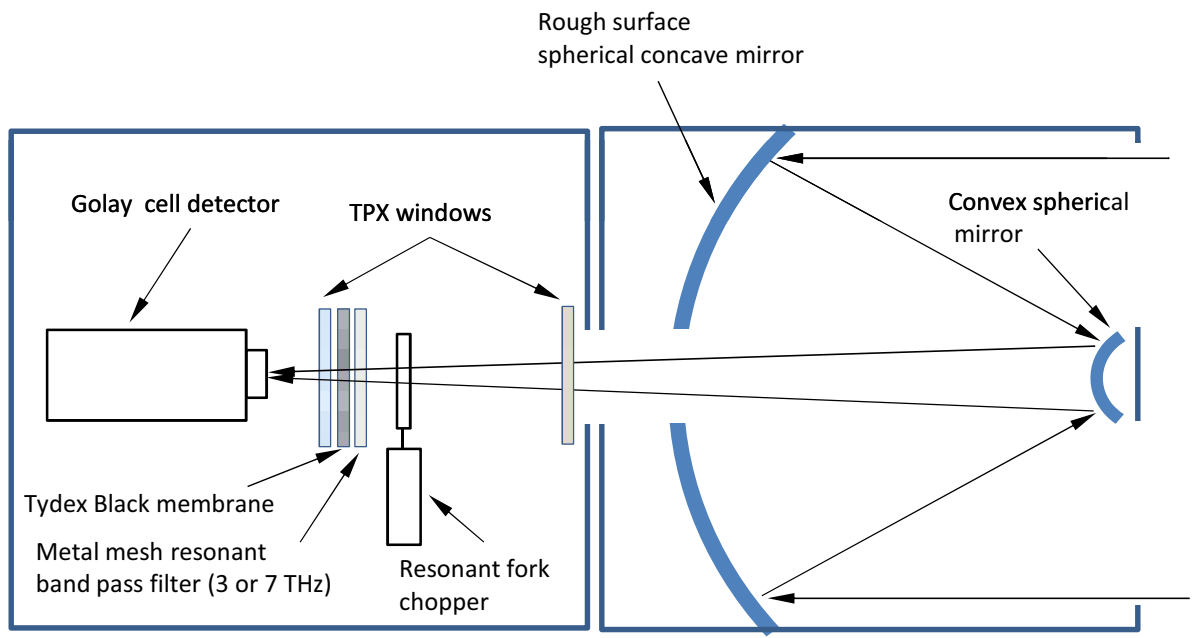

Figure 6

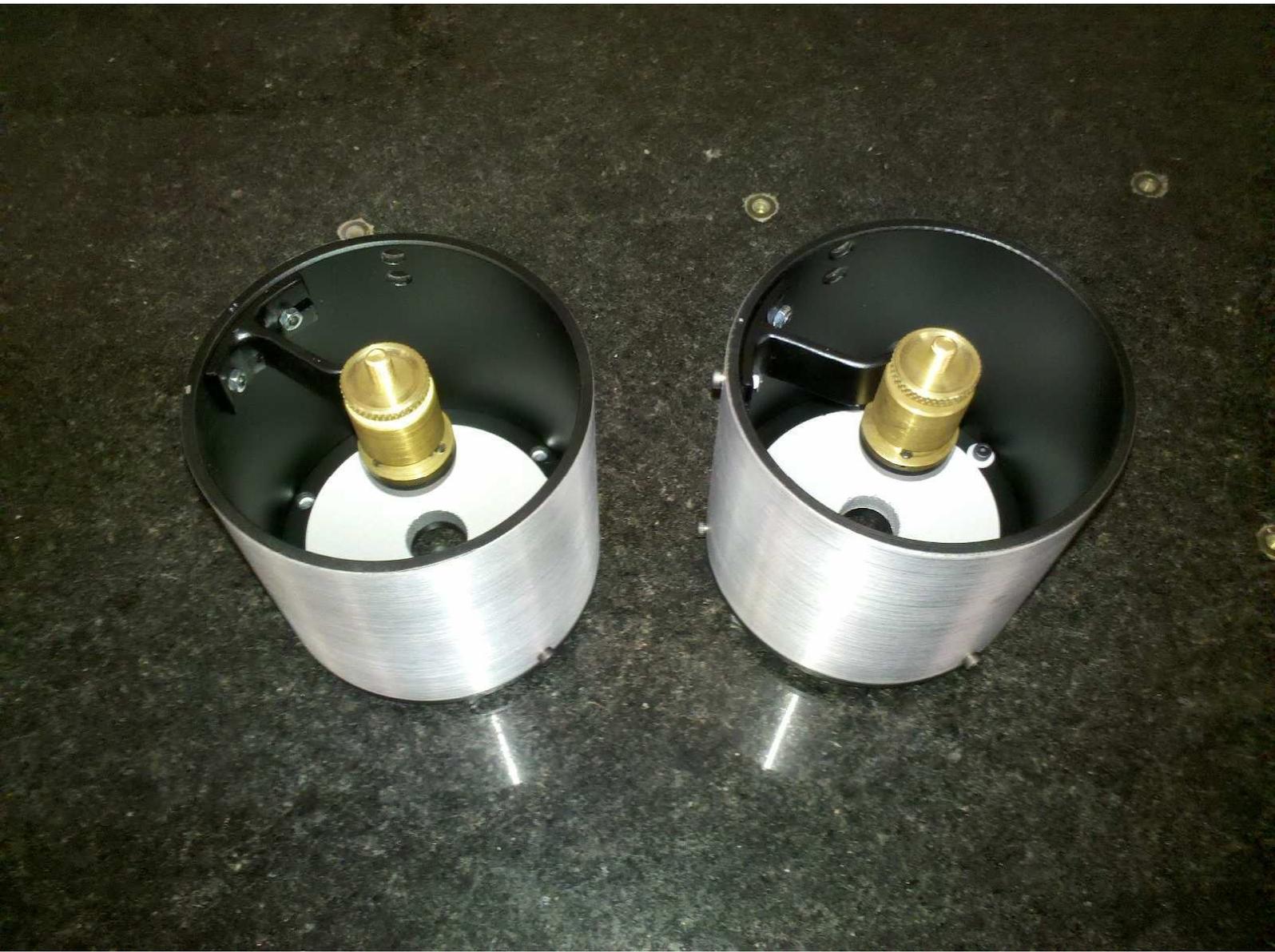

Figure 7

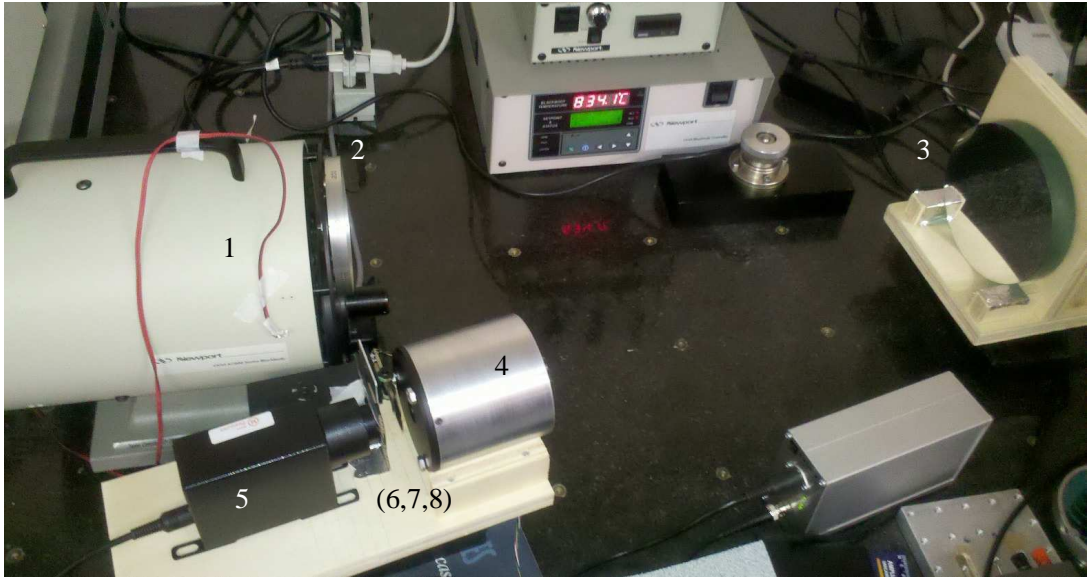

Figure 8